\documentclass[letterpaper,10pt]{article}
\usepackage{ictam}

\begin{document}

\title{UNSTEADY HEAT CONDUCTION PROCESSES IN A HARMONIC CRYSTAL WITH A SUBSTRATE POTENTIAL}

\author[1,2]{\underline{Mikhail B. Babenkov}
  \footnote{Email: mikhail.babenkov@gmail.com}}
\author[1,2]{Anton M. Krivtsov
  } 
\author[1,2]{Denis V. Tsvetkov
  }
\affil[1]{Peter the Great Saint Petersburg Polytechnic University, St. Petersburg, Russia}
\affil[2]{Institute for Problems in Mechanical Engineering RAS, St. Petersburg, Russia}

\maketitle

\begin{abstract}
An analytical model of high frequency oscillations of the kinetic and potential energies in a one-dimensional harmonic crystal with a substrate potential is obtained by introducing the nonlocal energies \cite{key01}. A generalization of the kinetic temperature (nonlocal temperature) is adopted to derive a closed equation determining the heat propagation processes in the harmonic crystal with a substrate potential.
\end{abstract}

\section{Equations of motion}
Non-Fourier heat conduction processes in ideal crystal structures have been intensively studied in the recent decades. The literature is surveyed in the review papers \cite{Bonetto,Lepri}. In this work we consider a one-dimensional crystal in the form of a chain of identical particles with mass $m$ that are connected by linear springs with each other and with a fixed base, described by the following equations of motion:
\begin{equation} \label{1}
    \ddot{u}_n=\omega^2_0 \big(u_{n-1}-\left(2+\epsilon \right)u_n+u_{n+1}\big)
,\qquad
    \epsilon = C_1/C_0, \qquad \omega_0=\sqrt{C_0/m}
\end{equation}
where
 $u_n$ is the displacement of the $n$-th particle,
 $m$ is the particle mass,
 $C_0$ is the stiffness of the interparticle bond,
 $C_1$ is the stiffness of the bond between a particle and the fixed base,
 and dots denote partial time derivatives.
The crystal is infinite: the index $n$ is an arbitrary integer.

The initial conditions \cite{key01,key02} are
\begin{equation} \label{2}
     u_n|_{t=0} = 0
     \qquad
     \dot u_n|_{t=0} = \sigma(x)\varrho_n,
\end{equation}
where $\varrho_n$ are independent random values with zero expectation and unit variance; $\sigma^2(x)$ is variance of the initial velocities, which is a slowly varying function of the spatial coordinate $x=na$, where~$a$ is the lattice constant. These initial conditions correspond to an instantaneous temperature perturbation, which can be induced in crystals, for example, by an ultrashort laser pulse.

\subsection{High frequency energy oscillations}
The dynamic transition of the kinetic energy into the potential energy of the bonds deformation is accompanied by high frequency oscillatory process with decreasing amplitude \cite{key03}. Similar oscillations can appear in solids during fast transient processes, for example, under the impact of a short laser impulse. In order to derive the equations describing such energy oscillations, we introduce the following nonlocal energies \cite{key01,key02}:
\begin{equation} \label{3}
K_n=\frac{1}{2}m \langle \dot{u}_{s} \dot{u}_{s+n} \rangle, \quad \mathit{\Pi}_n=\frac{1}{2} C_0 \langle \varepsilon_{s} \varepsilon_{s+n} \rangle + \frac{1}{2} C_1 \langle u_{s} u_{s+n} \rangle
,
\end{equation}
where $K_n$ is nonlocal kinetic energy,
$\mathit{\Pi}_n$ is nonlocal potential energy,
operator $\langle \dot u_{s} \dot u_{s+n} \rangle$ gives the covariance of the particles' velocities with the indexes difference of $n$,
$\varepsilon_{s}=u_{s}-u_{s-1}$ is the deformation of the bonds.
Formulae (\ref{3}) at $n = 0$ give the conventional energies.

Differentiation of the nonlocal energies with the use of dynamic equations (\ref{1}) allows us to derive the
following equation for the nonlocal Lagrangian $L_n$:
\begin{equation} \label{4}
\ddot{L}_n = 4 \omega^2_0 \big(L_{n-1}-\left(2+\epsilon\right)L_n+L_{n+1}\big), \qquad L_n=K_n-\mathit{\Pi}_n
\end{equation}
which coincides in form with the dynamic equation of chain (\ref{1}) and differs only by the value of the coefficient on the right side. Assuming that the initial velocities of various particles are independent and the initial displacements are absent, the initial conditions for (\ref{4}) can be written as \cite{key01}:
\begin{equation} \label{5}
L_n |_{t=0} = E \delta_n, \qquad \dot{L}_n |_{t=0} = 0,
\end{equation}
where $E$ is the full initial energy of the crystal, ${\delta_{n} = 1}$ for $n=0$ and ${\delta_{n} = 0}$ for $n\ne 0$.
The solution of the problem (\ref{4})--(\ref{5}) at $n = 0$ gives the conventional Lagrangian for the dynamical system (\ref{1}) in the form:
\begin{equation} \label{6}
L=E \left(J_0\left(2 \sqrt{4+\epsilon } \, \omega_0  t\right) -2 \sqrt{\epsilon } \,\omega_0 \int_0^t J_0 \left(2 \sqrt{4+\epsilon }\,  \omega_0 (t-\tau ) \right) J_1 \left(2 \sqrt{\epsilon }\, \omega_0  \tau \right) d\tau \right)
\end{equation}
Due to the complexity of this expression, two limiting cases are considered. The following asymptotic representation shows the behaviour of (\ref{6}) at the low values of $\epsilon$ (so-called ``soft substrate'' case, if $\epsilon\ll1$):
\begin{equation} \label{7}
L \simeq E\left(J_0\left(2 \sqrt{4+\epsilon }\, \omega_0  t \right)-\frac{1}{2}\sqrt{\epsilon } J_1\left(2 \sqrt{\epsilon } \,\omega_0  t \right)\right)
\end{equation}
Otherwise, if the value of $\epsilon$ is sufficiently high, then (\ref{6}) can be approximated as (``rigid substrate'' case, $\epsilon\gg1$):
\begin{equation} \label{8}
L \simeq E J_0 \Big( \left(\sqrt{4+\epsilon }-\sqrt{\epsilon }\right) \omega_0 t  \Big) \cos \Big( \left(\sqrt{4+\epsilon }+\sqrt{\epsilon }\right) \omega_0 t \Big)
\end{equation}
Considering the nonlocal energy conservation law \cite{key03}, one can obtain dependencies of the kinetic and potential energies on time: $$K(t)=E\,\frac{1+L(t)}2, \qquad \mathit{\Pi}(t)=E\,\frac{1-L(t)}2$$

\section{Nonlocal temperature}
We adopt an approach based on the covariance analysis \cite{Rieder,key02} for the velocities to obtain a closed equation system determining unsteady thermal processes. The nonlocal temperature $\theta_n(x)$ is introduced as \cite{key01,key02}:
\begin{equation} \label{9}
k_B(-1)^n \,\theta_n(x) = m\langle \dot u_i \dot u_j \rangle,
\end{equation}
where $k_B$ is the Boltzmann constant,
$n=j-i$ is the covariance index,
$x=\frac{i+j}2a$ is the spatial coordinate,
$a$ is the lattice constant.
If ${n=0}$ then $i=j$ and quantity $\theta_n$ coincides with the kinetic temperature $T$: $\theta_0(x,t) = T(x,t) = \frac m{k_B}\langle\dot u_i^2\rangle$, where $i=x/a$. The use of the correlation analysis \cite{key01,key02} and the long wavelength approximation allows one to obtain a differential-difference equation for $\theta_n$:
\begin{equation} \label{10}
\left(\theta_{n+1}+(2+\epsilon)\theta_{n}+\theta_{n-1}\right)\ddot{ }=-\frac{1}{4}c^{2}\left(\theta_{n+2}-2\theta_{n}+\theta_{n-2}\right)'',
\end{equation}
where primes denote partial coordinate derivatives.
The initial conditions for equation (\ref{10}) corresponding to original initial conditions (\ref{2}) are given by \cite{key01}:
\begin{equation} \label{11}
    \theta_n|_{t=0} = T_0(x)\delta_n \qquad
    \dot\theta_n|_{t=0} = 0 ,
\end{equation}
where
$T_0(x)=\frac1{2k_B}m\sigma^2(x)$
is the initial temperature distribution;
Problem (\ref{10})--(\ref{11}) can be solved by the means of discrete-time Fourier transform \cite{key02}, which allows to formulate the initial value problem for the kinetic temperature $T(x,t)$ in a simple form:
\begin{equation} \label{12}
    \ddot T + \frac1t\dot T = c_* ^2 T'', \qquad T|_{t=0} = T_0(x), \qquad \dot T|_{t=0} = 0,
\end{equation}
where $c_*=c \left(\sqrt{\epsilon +4}-\sqrt{\epsilon }\right)/2$ is the velocity of a heat wave propagating in a harmonic crystal with a substrate potential, $c=\omega_0 a$ is the sound velocity in a simple harmonic crystal.

\end{document}